\documentclass[10pt]{article}
\newcommand{\version}{\today}

\usepackage{amsthm,amsfonts,amsmath, amscd}

\swapnumbers
                              
\theoremstyle{plain}
\newtheorem{thm}{THEOREM}[section]

\theoremstyle{definition}

\newtheorem{remark}[thm]{Remark}
\theoremstyle{remark}
\newcommand{\upchi}{\raise1pt\hbox{$\chi$}}

\newcommand{\hn}{{\mathord{\widehat{n}}}}

\newcommand{\tr}{{\rm Tr}}
\renewcommand{\|}{{\Vert}}
\numberwithin{equation}{section}
\newcommand{\ef}{{\rm E}_{\rm f}}
\newcommand{\sef}{{\rm E}_{\rm sq}}

\newcommand{\un}{{\rm 1\kern -2.5pt l}}

\def\Tr{{\rm Tr}}
\begin{document}
%%%%%%%%DRAFT%%%%%%%
\iffalse
\fi
%%%%%%%%%%%%%%%%%%%%%%
\markboth{\scriptsize{CL \version}}{\scriptsize{CL March 27, 2012}}
\def\mn{{\bf M}_n}
\def\hn{{\bf H}_n}
\def\hnp{{\bf H}_n^+}
\def\hmnp{{\bf H}_{mn}^+}
\def\h{{\cal H}}
\title{{\sc  Bounds for  
Entanglement  via an Extension of Strong
Subadditivity of Entropy}}
\author{
\vspace{5pt}  Eric A. Carlen$^{1}$  and Elliott H. Lieb$^{2}$ \\
\vspace{5pt}\small{$^{1}$ Department of Mathematics, Hill Center,}\\[-6pt]
\small{Rutgers University,
110 Frelinghuysen Road
Piscataway NJ 08854-8019 USA}\\
\vspace{5pt}\small{$^{2}$ Departments of Mathematics and Physics, Jadwin
Hall, Princeton University}\\[-6pt]
\small{ P.~O.~Box 708, Princeton, NJ  08542.}\\[-6pt]
 }

\date{March 27, 2012}
\maketitle

\centerline{\it We dedicate this paper to Walter Thirring
on the occasion of his 85th birthday}

\footnotetext [1]{Work partially supported by U.S.
National Science Foundation grant  DMS 0901632.}

\footnotetext [2]{Work partially supported by U.S. National
Science Foundation
grant PHY 0965859 \hfill\break
\copyright\, 2012 by the authors. This paper may be reproduced, in its
entirety, for non-commercial purposes.
}

\begin{abstract}
Let $\rho_{12}$ be a bipartite density matrix. We prove
lower bounds for the entanglement of formation 
$\ef(\rho_{12})$ and  the squashed entanglement
$\sef(\rho_{12})$ in terms of the conditional entropy
$S_{12} - S_1$, and prove that these
bounds are sharp by constructing a new class of states whose
entanglements can be computed, and for  which the bounds
are saturated.
\end{abstract}

\medskip
\centerline{Key Words: entropy, entanglement, strong subadditivity.}

%%%%%%%%%%%%%%%%%%%%%%%%%%%%%%%%%%%%%%%%%%%%%%%
%%%%%%%%%%%%%%%%%%%%%%%%%%%%%%%%%%%%%%%%%%%%%%%
\section{Introduction}
%%%%%%%%%%%%%%%%%%%%%%%%%%%%%%%%%%%%%%%%%%%%%%%

\maketitle

\def\HH{{\mathcal{H}}}

It is well known that in  classical probability theory
the entropy of a bipartite density,
$\rho_{12}$, and its marginal densities $\rho_1$ and
$\rho_2$, satisfy the {\it positivity of the  conditional
entropies:}
\begin{equation} \label{conditional}
S_{12} - S_1 \geq 0 
\qquad {\rm and} \qquad  S_{12} - S_2\geq 0 \ .
\end{equation}
In quantum mechanics one has a density matrix $\rho_{12}$
and its reduced density matrices $\rho_1$ and
$\rho_2$, for which one can define von Neumann entropies,
but the analog of \eqref{conditional} need
not hold. The 
closest one can come to \eqref{conditional} for the von
Neumann entropy is the triangle inequality \cite{AL},
$S_{12} - |S_1 - S_2| \geq 0$

The main thrust of our paper is to  show that the failure of
either one  of the inequalities \eqref{conditional}, which can
occur  only in quantum mechanics,  necessarily implies
quantum entanglement, which is a peculiar correlation found
only in quantum mechanics.
More precisely, our main result will be the sharp inequality
\begin{equation*}
{\rm E} \geq \max \{S_1-S_{12}, \ S_2-S_{12}, \ 
0 \}
\end{equation*}
where ${\rm E}$ denotes either one of two  measures of
entanglement,
$\ef$ and $\sef$, which we define now. 

Let $\rho_{12}$ be a density matrix on a bipartite system; i.e.,
on a tensor product of two Hilbert spaces
$\HH_1\otimes \HH_2$.  Then $\rho_{12}$ is {\em finitely
separable}
in case it can be decomposed as a convex combination of
tensor products:
\begin{equation}\label{sep}
\rho_{12} = \sum_{k=1}^n \nu_k \rho_1^k \otimes \rho_2^k
\end{equation}
where the $\nu_k$ are positive and sum to $1$, and each
$\rho_\alpha^k$ is a density matrix on $\HH_\alpha$.  A
bipartite state is {\em separable} if it is in the closure
of the set of finitely separable states.
A bipartite state that is not separable is 
{\em entangled}.  

The first measure of entanglement that we study is the 
{\em
entanglement of formation} $\ef$, introduced by Bennett
{\it et al.} \cite{B1,B2}, which
is defined in terms of the von Neumann entropy 
$S(\rho) = -\Tr(\rho  \log \rho)$ by
\begin{equation} \label{formation}
\ef(\rho_{12}) = \inf \left\{ \sum_{j=1}^n \lambda_j 
S(\tr_2 \omega_j) \ :\ \rho_{12} = \sum_{k=1}^n\lambda_k
\omega^k \ \right\}
\end{equation}
where $\tr$ and $\tr_2$, respectively, are the trace over
the tensor product $\HH_1\otimes \HH_2$ and the partial 
trace over $\HH_2 $ alone. 
The coefficients
$\lambda_k$ in the expansion are required to be positive and
sum to $1$, and each $\omega^k$ is a state on $\HH_1\otimes
\HH_2$, which, by the concavity of $S$, may be taken to be a
pure
state without affecting the value of the infimum.  Since the two partial traces of a 
pure state have the same spectrum and hence the same entropies \cite{AL},
$\ef(\rho_{12})$ is symmetric in $1$ and $2$.
It is
known that $\ef(\rho_{12}) =0$ if and only if $\rho_{12}$ is
separable; see \cite{BCY} for a discussion of this result in relation to other measures of entanglement.
A variational problem for quantum entropy in a general von Neumann 
algebra setting that is  related to (\ref{formation}) was
introduced by Narnhofer and Thirring \cite{NT1}.

The second measure of entanglement that we study
is a smaller quantity, the {\em
squashed entanglement}, which was first defined by Tucci
\cite{T}. It was rediscovered
by Christandl
and Winter \cite{CW1} who showed that it has many important
properties, such as additivity. 
In several ways it provides a better estimate of the purely
quantum mechanical
entanglement than entanglement of formation. For a review
of the subject, see \cite{H4}.

 The definition of squashed entanglement involves a
relaxation of the variational
problem defining
$\ef$; i.e., it extends  the domain over which the
minimization is to be taken, in the following way:

Given a decomposition of a bipartite state ${\displaystyle
\rho_{12} =  \sum_{k=1}^n\lambda_k \omega^k}$, we may
associate a tripartite state $\rho_{123}$
on $\HH_1\otimes \HH_2\otimes \HH_3$ where $\HH_3$ is
{\it any } Hilbert space of dimension at least $n$ by
letting$ \{\phi_1,\dots,\phi_n\}$ be orthonormal in
$\HH_3$, and defining
\begin{equation}\label{particular}
\rho_{123} = \sum_{k=1q}^n  \lambda_k \omega^k \otimes
|\phi_k\rangle\langle \phi_k|\ .
\end{equation}
Using only the fact that each $\phi_j$ is a unit vector,
and not  using the orthogonality, one has that
$\rho_{12} = \Tr_3\left(\rho_{123}\right)$, so that
$\rho_{123}$
is an {\em extension} of $\rho_{12}$, meaning that $\tr_3\rho_{123} = \rho_{12}$.  (This is
only one of many ways one could extend $\rho_{12}$ to a tripartite state $\rho_{123}$. Another way is
through {\em purification} which we shall use in Section 2. The main point to note at present is that there are infinitely many extensions.)

Let $\rho_{23}$
denote $\tr_{1}(\rho_{123})$, 
let $\rho_{3}$ denote $\tr_{12}(\rho_{123})$,  let
$S_{23}$ denote $S(\rho_{23})$, and let $S_3$ denote $S(\rho_3)$,
etc. following the notational scheme in \cite{L}.
A simple computation shows that for the extension $\rho_{123}$ given in (\ref{particular}),  now using the fact that
$ \{\phi_1,\dots,\phi_n\}$ is orthonormal, 
\begin{equation}\label{PIV}
S_{13}+S_{23} - S_{123} - S_3 = 2  \sum_{j=1}^n \lambda_j  S(\tr_2 \omega_j)\ .
\end{equation}
The right side is (twice) the quantity appearing in the definition of $\ef$, (\ref{formation}).

We  recall a standard definition: 
For {\em any} density matrix $\rho_{123}$ on $\HH_1\otimes\HH_2\otimes\HH_3$,   
the {\em conditional mutual information of $1$ and $2$ given $3$} is the quantity $I(1,2|3)$
defined by
\begin{equation}\label{CMI}
I(1,2|3) := S_{13}+S_{23} - S_{123} - S_3 \geq 0\ .
\end{equation}
The   {\em strong subadditivity of the quantum entropy}
theorem of   Lieb and Ruskai \cite{LR}, is the statement
that for all tripartite states $\rho_{123}$, 
\begin{equation}\label{SSA}
I(1,2|3)  \geq 0\ .
\end{equation}
By these results and (\ref{PIV}),  the quantity $\sef(\rho_{12})$ defined by
\begin{equation}\label{squashed}
\sef(\rho_{12}) =  \frac12 \inf\{ \ I(1,2|3) \ : 
\rho_{123} \ {\rm is}\ any\ {\rm tripartite\ extension\ of }\
\rho_{12}\ \}\ .
\end{equation}
defines a non-negative minorant to $\ef(\rho_{12})$ known as the
{\em squashed entanglement} of a bipartite
state $\rho_{12}$,

Evidently, for all $\rho_{12}$, 
$\ef(\rho_{12}) \geq \sef(\rho_{12})$.  Moreover, as proved in \cite{HJPW}, there is
equality in  (\ref{SSA}) if and only if
$\HH_3$ has the form 
$$\HH_3 = \bigoplus_{j=1}^m \HH^j_{3\ell}\otimes \HH_{3r}^j$$ and 
$\rho_{123}$ has the form
\begin{equation}\label{eqssa}
\rho_{123} = \bigoplus_{j=1}^m \rho_{1,3\ell}^j\otimes \rho_{3r,2}^j\ . 
\end{equation}
Evidently, for any $\rho_{123}$ of the form (\ref{eqssa}),
$\rho_{12} := \tr_2(\rho_{123})$ is separable. Moreover, if
$\rho_{12}$ is separable, and has the decomposition
(\ref{sep}),
and if we take $\rho^j_{1,3\ell}$ to be an
arbitrary
purification of $\rho_1^j$ onto $\HH_1\otimes
\HH_{3\ell}^j$, we can take $\HH_{3r}$ to be
one-dimensional, 
$\rho_{3r,2}^j = \rho_2^j$, and with these definitions,
$\rho_{123}$ is an extension of the given separable
bipartite state $\rho_{12}$ for which equality holds in 
(\ref{SSA}).  Thus, whenever $\rho_{12}$ is finitely 
separable, then $\sef(\rho_{12}) =0$. It is elementary to
see that 
whenever $\rho_{12}$ is finitely separable, then
$\ef(\rho_{12}) =0$.  
A  continuity argument \cite{AF} then shows that $\ef(\rho_{12})
=0$, and hence $\sef(\rho_{12}) =0$,
whenever $\rho_{12}$ is separable. The converse is not obvious, but it has 
recently been proved in \cite{BCY} that whenever $ S_{13}+S_{23} - S_{123} - S_3$ is sufficiently small, 
then $\rho_{123}$ is well -approximated by a density $\rho_{123}$ that has 
the form (\ref{eqssa}).  Thus, $\sef(\rho_{12}) >0$, and hence $\ef(\rho_{12})>0$, whenever $\rho_{12}$
is not separable. Hence both $\ef$  and $\sef$ are {\em faithful} measures of entanglement. 

In general, it is not a simple matter to evaluate the infima that define  $\ef$  
and $\sef$. In particular, even if the bipartite state $\rho_{12}$ operates on 
a finite dimensional Hilbert space, there is no known {\em a-priori} bound on the dimension 
of the Hilbert spaces $\HH_3$ that must be used to nearly minimize $I(1,2|3)$. 
This makes $\sef$ difficult to evaluate in general, so that sharp lower bounds are of interest.

The non-negativity of $\sef$, as we have explained, is a direct consequence of the strong 
subadditivity inequality (\ref{SSA}).  We shall  prove an extension of (\ref{SSA}), and show that it provides
sharp lower bounds on $\ef$ and $\sef$. 

\begin{thm}[Extended strong subadditivity]\label{essa}
For all tripartite states $\rho_{123}$,
\begin{equation}\label{essa1}
I(1,2|3) \geq 2 \max\{ S_1 - S_{12} \ ,  S_2 - S_{12}\
,  0\ \} \ .
\end{equation}
The inequality $I(1,2|3) \geq \lambda \max\{ S_1 -
S_{12} \ ,  S_2 - S_{12}\ ,  0\ \} $ can be violated for all
$\lambda>2$.
\end{thm}

The inequality (\ref{essa1}) extends the inequality
(\ref{SSA}) in an obvious way, but as we shall see, it is
actually a {\em consequence} of the seemingly weaker
inequality (\ref{SSA}).
We prove Theorem~\ref{essa}
in Section 2.

As a direct consequence of Theorem~\ref{essa}, we obtain lower bounds for $\ef$ and $\sef$ that we shall show to be sharp:

\begin{thm}[Lower bounds for  $\sef$ and $\ef$]\label{lbef}
For all bipartite states $\rho_{12}$,
\begin{equation}\label{lbesf1}
\sef(\rho_{12}) \geq \max\{ S_1 - S_{12} \ ,\  S_2 - S_{12}  \ ,\ 0\} \ .
\end{equation}
and
\begin{equation}\label{lbef2}
\ef(\rho_{12}) \geq \max\{ S_1 - S_{12} \ ,\  S_2 - S_{12}  \ ,\ 0\} \ .
\end{equation}
Both of these inequalities are sharp in that there  exists a class of  bipartite states $\rho_{12}$ for which
\begin{equation}\label{lbef3}
\ef(\rho_{12}) = \sef(\rho_{12}) =  S_1 - S_{12} > 0\ ,
\end{equation}
and for which $S_1$ and $S_{12}$ may take arbitrary non-negative values.
\end{thm}

Note that because $\ef \geq \sef$,  the inequality
(\ref{lbef2}) is implied by (\ref{lbesf1}), whereas the fact that  (\ref{lbef2}) is sharp implies that  (\ref{lbesf1}) is sharp.

A weaker form of the inequality (\ref{lbesf1}) was given by
Christandl and Winter
\cite{CW1}.
Their lower bound involves the
averaged  quantity 
\begin{equation}\label{wkr}
\frac12 (S_1+S_2) -S_{12}
\end{equation}
 in place of
$\max\{S_1-S_{12}, S_2-S_{12}\}$. The difference can be significant: As we explain in Remark~\ref{size} below, 
there exist states $\rho_{12}$  for which $\sef(\rho_{12}) = S_1 - S_{12}$ is 
arbitrarily large, but the quantity in (\ref{wkr}) is negative. 
Moreover, the argument in \cite{CW1}
relied on a lower bound for the  the {\em one-way
distillable entanglement} 
in terms of the conditional entropy. This inequality, known
as the {\em hashing inequality}  had been a long-standing
conjecture,
and its proofs remain complicated.  Our contribution is to
show how this stonger lower bound follows in a relatively
simple manner from strong subadditivity, and to provide the
examples that prove the
sharpness of these bounds.

. 

The class of bipartite states referred to in the final part
of
Theorem~\ref{lbef}  are states that saturate the Araki-Lieb
triangle
inequality \cite{AL} 
\begin{equation}\label{ALI}
S_{12} \geq |S_1 - S_2|\ .
\end{equation}
A characterization of cases of equality has been known
for some time, and discussed as an exercise in \cite[Ex.
11.16]{NC}. Recently, Zhang and Wu \cite{ZW} proved
this characterization by using
the conditions for equality in the more difficult
strong subadditivity theorem proved by Hayden {\it et al.}
\cite{HJPW}.
We give a
short and elementary proof which provides a more detailed
characterization of the cases of equality.

\begin{thm}[Bipartite states with $S_{12} = S_1 - S_2$]\label{sharp}  A  bipartite state  $\rho_{12}$ satisfies  
$$S_{12} = S_1 - S_2$$
if and only if
\begin{equation}\label{rank1}
{\rm rank}(\rho_1) =   {\rm rank}(\rho_2) {\rm rank}(\rho_{12})
\end{equation}
and  $\rho_{12}$ has a spectral decomposition of the form
\begin{equation}\label{rank2}
\rho_{12} = \sum_{j=1}^{ {\rm rank}(\rho_{12})}\kappa_j |\phi_j\rangle\langle \phi_j|\ ,
\end{equation}
where for each $i,j$, 
\begin{equation}\label{rank3}
\tr_1 |\phi_i\rangle\langle \phi_j| = \delta_{i,j}\rho_2\ .
\end{equation}

Examples of such states may be constructed by choosing any set $n$ of non-negative 
numbers $\kappa_j$ with $\sum_{j=1}^n\kappa_j =1$, and any state $\rho_2$ on $\HH_2$, and taking
$\sigma_j$ to be a purification of $\rho_2$ on $\HH_1\otimes \HH_2$ such that the ranges of the $\sigma_j$ 
are mutually orthogonal (which is possible under condition (\ref{rank1})).
By  defining
$$\rho_{12} = \sum_{j=1}^m \kappa_j \sigma^j$$
one has
$$S_{12} = -\sum_{j=1}^n\kappa_j\log\kappa_j \qquad S_2 = S(\rho_2) \qquad{\rm and}\qquad S_1 = S_{12} + S_2$$
so that the construction yields examples in which $S_2$ and $S_{12}$ take  arbitrary non-negative values. 

Moreover, for any  $\rho_{12}$ yielding equality in the triangle inequality, 
\begin{equation}\label{iden}
\sef(\rho_{12}) = \ef(\rho_{12}) = S_1 - S_{12}\ .
\end{equation}
\end{thm}

\begin{remark}\label{size}
Theorem~\ref{sharp} yields bipartite states $\rho_{12}$  for which  $S_1 - S_{12} = S_2$, 
and for which $S_1$ and $S_{12}$, and   can have  arbitrary non-negative values. 
Therefore,  we can arrange that $S_! - S_{12} = S_2$
is positive, and even arbitrarily  large, but the quantity in (\ref{wkr}) is negative.    

For our immediate purpose here,  we do not require the full strength of 
Theorem~\ref{sharp} which characterizes {\em all} bipartite states $\rho_{12}$  with 
equality in the triangle inequality $S_{12} \geq |S_1 - S_2|$.  To show that our bounds in 
Theorems~\ref{essa} and \ref{lbef} are sharp, it suffices to observe that the construction described in Theorem~\ref{sharp}
yields {\em examples} of  bipartite states $\rho_{12}$  for which  $S_1 - S_{12} = S_2$, 
and for which $S_1$ and $S_{12}$, and   can have  arbitrary
non-negative values, and this is an easy calculation.
\end{remark}

An {\it uppper bound} for $\ef $, and hence for $\sef$ as
well, is known under  the rubric {\it entanglement never
exceeds local entropy}.
If one  inserts the trivial decomposition
$\rho_{12}= \rho_{12} $ into the basic definition
\eqref{formation}, and symmetrizing the bound, one obtains
the upper bound:
\begin{equation}
\ef(\rho_{12}) \leq \min\{S_1,\ S_2\}\ .
\end{equation}
For the states used in \eqref{iden}, we see that this upper
bound is sharp for  both $\ef$ and $\sef$.  We are grateful
to Matthias Christandl for pointing this out to us.

\section{Proofs}

We shall use {\em purification} arguments, which appears to
have been first  used in \cite{AL} to prove the Araki-Lieb
triangle inequality. In Lemmas~3 of \cite{AL} it is shown
that for any pure bipartite state $\rho_{12}$,
$\rho_1$ and $\rho_2$  have the same non-zero spectrum, and hence the same entropy. 
Lemma~4  of \cite{AL}  shows that given any density matrix $\rho$ on $\HH$, there is a pure state
$\rho_{12}$ on $\HH\otimes \HH$ such that $\tr_2(\rho_{12}) = \rho$. 

Using these lemmas, the triangle inequality may then be
deduced from the subadditivity of the entropy; i.e., the
inequality
$S_{23} \leq  S_2+S_3$. Considering any purification $\rho_{123}$ of $\rho_{23}$, $S_{23} = S_1$ and $S_{3} = S_{12}$ and hence
$
S_{12} \geq S_1 - S_2$.
By symmetry, one then has (\ref{ALI}).  (Note that one can use essentially the same purification to recover the 
subadditivity inequality from the triangle inequality; in this sense the inequalities are equivalent.)

\medskip

\noindent{\bf Proof of Theorem~\ref{essa}:}  The starting
point of our proof is inequality \eqref{essa0} below, which
appears in \cite[Theorem 2]{LR}, and whose simple proof we
recall:
Consider any purification $\rho_{1234}$ of $\rho_{123}$.
Then since $\rho_{1234}$ is pure,   $S_{23} = S_{14}$ and
$S_{123} = S_4$
Then
$$S_{12} + S_{23} - S_1 - S_3 = S_{12} + S_{14} - S_{124}-S_1\ ,$$
and the right hand side is non-negative by (\ref{SSA}). 
This proves 
\begin{equation}\label{essa0}
S_{12}+S_{23} \geq S_1+S_3\ .
\end{equation}

Next, adding  $S_{12}+S_{23} \geq S_1+S_3$ and   $S_{13}+S_{23} \geq S_1+S_2$, we obtain
\begin{equation}\label{essa00}
 S_{12} + S_{13} + 2S_{23}  \geq 2S_1 + S_2 + S_3 \ .
 \end{equation}
Consider any purification $\rho_{1234}$ of $\rho_{123}$.  Then we obtain, using $S_{12} = S_{34}$,
$S_{23} = S_{14}$, and $S_2 = S_{134}$,
$$
S_{13}+S_{34} -S_{134} -S_3 \geq 2(S_{1} - S_{14})\ ,
$$
which is (\ref{essa1}) with different indices.   As a consequence of Theorem~\ref{sharp},  (\ref{essa1}) is sharp. 
\qed

\medskip

\begin{remark}  The inequality (\ref{essa00}) is the crux of the matter. In a similar way, one may deduce the related inequalities
\begin{equation}\label{essa0B}
2S_{12} + S_{13} + 2S_{23}  \geq 2S_1 + S_2 + 2S_3  
\qquad{\rm and}\qquad S_{12} + S_{13} + S_{23}  \geq  S_1 + S_2 + S_3\ .
\end{equation}
The second inequality in (\ref{essa0B}) is obtained by averaging (\ref{essa0}) 
over permutations of the indices. One may then add (\ref{essa0}) to this to obtain the first inequality.
It is worth noting that the purification argument in leading from strong subadditivity to  
(\ref{essa0}) reverses, so that (\ref{essa0}) is equivalent to strong subadditivity. 
\end{remark}
\medskip

\noindent{\bf Proof of Theorem~\ref{lbef}:}   The inequality (\ref{lbef2}) is an immediate consequence of 
(\ref{essa1}) and the definition of $\sef$. Then (\ref{lbef2}) follows since $\ef \geq \sef$. 
Once more, the statement about sharpness is a consequence of Theorem~\ref{sharp}. \qed

\begin{remark} It is worth pointing out that the inequality
(\ref{lbef2}), but not (\ref{lbesf1}),  has a direct proof using the concavity of the
conditional entropy  \cite[Theorem 1]{LR}, which may be seen as a consequence of 
the joint convexity of the relative entropy theorem: The relative
entropy 
of two states $\rho$ and $\sigma$ is defined to be
$S(\rho|\sigma) := \tr[\rho(\log \rho - \log \sigma)]$,
which is jointly convex as a limiting case of  the
concavity theorem \cite[Theorem 1]{L}. Then
defining $\sigma_2$ to be the normalized identity on
$\HH_2$, 
$$S(\rho_{12}|\rho_1\otimes \sigma_2)  = \tr \rho_{12}\log \rho_{12} - \tr \rho_1\log \rho_1  -\log({\rm dim}(\HH_2))\ .$$

By the concavity of $\rho_{12}\mapsto S_{12} - S_2$,  for any decomposition of 
$\rho_{12}$ of the form ${\displaystyle  \sum_{k=1}^n\lambda_k \omega^k }$, in which the $\omega_k$ are pure,
$$S_{12} - S_1 \geq    \sum_{k=1}^n\lambda_k [S(\omega^k ) - S(\tr_2\omega^k )] = - \sum_{k=1}^n\lambda_kS(\tr_2\omega^k )]$$
since $S(\omega^k) =0$.
\end{remark}

\medskip
\noindent{\bf Proof of Theorem~\ref{sharp}:}   As we have explained, the inequality 
$S_{12} \geq |S_1- S_2|$ is deduced from the subadditivity inequality $S_{23} \leq S_2+S_3$.
Consequently, if $\rho_{12}$ is any bipartite state for which $S_{12} = S_1 - S_2$, 
and $\rho_{123}$ is any purification of it,  $S_{23} = S_2+S_3$ and hence
 \begin{equation}\label{ssa3}
 \rho_{23} = \rho_{2}\otimes\rho_{3}
 \end{equation}
since only product state saturate the subadditivity inequality.   

Next,  define 
$d_j := {\rm rank}(\rho_j)$ for $j=1,2,3$, 
Likewise, define
$$d_{12} := {\rm rank}(\rho_{12})\quad  \ ,\quad   d_{23} := {\rm rank}(\rho_{23}) \ {\rm etc}. $$

Since $\rho_{123}$ is pure,
$d_{23}  = d_1$ and $ d_{12} = d_3$.
But by (\ref{ssa3}),
$d_{23} =    d_2d_3$, and
therefore,
 \begin{equation}\label{ssa3A}
  d_1 = d_2 d_{12}\ ,
  \end{equation}
  which proves the necessity of (\ref{rank1}).
  
We now show that if $S_{12} = S_1 - S_2$, then $\rho_{12}$ has a {\em spectral decomposition} of the from (\ref{rank2})
where, for each $i,j$, (\ref{rank3}) is satisfied. 
This will prove that every bipartite state $\rho_{12}$ for which  
$S_{12} = S_1 - S_2$ has the structure asserted in Theorem~\ref{sharp}. 

Conversely,  when $\rho_{12}$  has the spectral projection  (\ref{rank2}),  
$S_2 = -\kappa_j \log \kappa_j$. Moreover, when (\ref{rank3}) is satisfied,
the $\tr_2|\phi_j\rangle\langle\phi_j|$ have mutually orthogonal ranges, and each has entropy $S_2$, and so
$$
S_1 = -\sum_{j=1}^{ {\rm rank}(\rho_{12})} {\kappa_j}\log\kappa_j + 
\sum_{j=1}^{ {\rm rank}(\rho_{12})} \kappa_j S_2  = S_{12} + S_2\ .
$$ 
Thus,  every 
every bipartite state with the structure described in (\ref{rank2}) and (\ref{rank3})  satisfies $S_{12} = S_1 - S_2$.

To prove that the spectral decomposition (\ref{rank2}) of $\rho_{12}$  satisfies  (\ref{rank3}), 
let  $\varphi \in \HH_1\otimes \HH_2\otimes \HH_3$ be such that $\tr_3|\varphi\rangle\langle \varphi| = \rho_{12}$. 
Note that without loss of generality, we may assume that ${\rm dim}(\HH_j) = d_j$ for $j=1,2,3$. 
Let us choose bases for $\HH_2$ and $\HH_3$ in which $\rho_2$ and $\rho_3$ are diagonal, and pick any orthonormal basis for $\HH_1$.
For $j=1,2,3$, let $X_j$ be a set of cardinality $d_j = {\rm dim}(\HH_j)$. Using the orthonormal bases selected above, we may view the vector
$\varphi \in \HH_1\otimes \HH_2\otimes \HH_3$ with a function $\varphi(x_1,x_,x_3)$ on $X_!\times X_2\times X_3$. 

Then the pure state density matrix $\rho_{123}$ has the matrix elements
$$\varphi(x_1,x_2,x_3)\varphi^*(x_1',x_2',x_3')\  .$$

Taking partial traces, and using the fact that our bases diagonalize $\rho_2$ and $\rho_3$
$$\rho_{23}(x_2,x_3;x_2',x_3')  = \sum_{x_1}\varphi(x_1,x_2,x_3)\varphi^*(x_1,x_2',x_3') \ ,$$
$$\rho_{2}(x_2,;x_2')  = \sum_{x_1,x_3}\varphi(x_1,x_2,x_3)\varphi^*(x_1,x_2',x_3)  = \lambda_{x_2}\delta_{x_2,x_2'}\ ,$$
and
 \begin{equation}\label{ssa4}
 \rho_{3}(x_3;,x_3')  = \sum_{x_1,x_2}\varphi(x_1,x_2,x_3)\varphi^*(x_1,x_2,x_3') = \mu_{x_3}\delta_{x_3,x_3'}\ .
 \end{equation}
Then  by (\ref{ssa3}), 
 \begin{equation}\label{ssa5}
\sum_{x_1}\varphi(x_1,x_2,x_3)\varphi^*(x_1,x_2',x_3')   =  \lambda_{x_2}\mu_{x_3}\delta_{x_2,x_2'}\delta_{x_3,x_3'} \  .
\end{equation}
Now for each $x_2\in X_2$, $x_3\in X_3$, define $\psi_{x_2,x_3}(x_1)$ by
$\psi_{x_2,x_3}(x_1) = \varphi(x_1,x_2,x_3)$.
It follows from (\ref{ssa5}) that
 \begin{equation}\label{ssa8}
 \{ \psi_{x_2,x_3} \ :\ x_2\in X_2,x_3\in X_3\}
 \end{equation}
is an orthogonal set of vectors. Moreover, since by assumption each 
$\lambda_{x_2}$ and each $\mu_{x_3}$ is non-zero, none of these vectors is zero:
The set of vectors in (\ref{ssa8}) is an orthogonal basis for $\HH_1$. 

In the same way, defining
$\chi_{x_3}(x_1,x_2) := \varphi(x_1,x_2,x_3)$,
 \begin{equation}\label{ssa9}
 \{ \chi_{x_3} \ : \   x_3\in X_3\}
 \end{equation}
 is a set of $d_3$ orthogonal vectors in $\HH_1\otimes\HH_2$, and by (\ref{ssa4}), $\|\chi_{x_3}\|^2 = \mu_{x_3}$. Now define
 $$\eta_{x_3} := \frac{1}{\sqrt{\mu_{x_3}}}\chi_{x_3}\ .$$
  Then since 
 ${\displaystyle \rho_{12}(x_1,x_2;x_1',x_2')  = \sum_{x_3}\varphi(x_1,x_2,x_3)\varphi^*(x_1',x_2',x_3)}$
 $$\rho_{12} = \sum_{x_3} \mu_{x_3}|\eta_{x_3}\rangle \langle \eta_{x_3}|\ .$$

 Now we may rewrite
 $$\sum_{x_1}\varphi(x_1,x_2,x_3)\varphi^*(x_1,x_2',x_3')    = 
 \sum_{x_1} \sqrt{\mu_{x_3}}  \sqrt{\mu_{x_3'}} \eta_{x_3}(x_1,x_2) \eta^*_{x_3'}(x_1,x_2')\ .$$
 Comparing with (\ref{ssa5}), we see that 
 $$\sum_{x_1} \sqrt{\mu_{x_3}}  \sqrt{\mu_{x_3'}} \eta_{x_3}(x_1,x_2) \eta^*_{x_3'}(x_1,x_2')  =  
 \lambda_{x_2}\mu_{x_3}\delta_{x_2,x_2'}\delta_{x_3,x_3'} \  ,$$
 and this proves (\ref{rank3}).

We now prove  the final statement.    Let $\rho_{12}$ be such that $S_{12} = S_1- S_2$. Then 
${\displaystyle \rho_{12} = \sum_{j=1}^m \lambda_j \sigma^j}$
where each $\sigma^j$ is a purification of $\rho_2$. Hence
$$\ef(\rho_{12}) \leq \sum_{j=1}^m \lambda_j S(\tr_1(\sigma^j) =  \sum_{j=1}^m \lambda_j S_2 = S_2 = S_1 - S_{12}\ .$$
By the lower bound in Theorem~\ref{lbef}, we must have $\ef(\rho_{12}) = 
\sef(\rho_{12}) = S_1 -S_{12}$.
\qed

\bigskip

\noindent {\bf Acknowledgement:}
We thank Heide
Narnhofer and Walter Thirring for
sharing with us a draft version of their review article
\cite{NT}, which
inspired us to consider the problems addressed here.

 \end{document}